\documentstyle[11pt,paspconf]{article}

\def\plotfiddle#1#2#3#4#5#6#7{\centering \leavevmode
\vbox to#2{\rule{0pt}{#2}}
\includegraphics{#1}}
\def\msun{{\rm M}_\odot}

\begin{document}

\title{\vspace{-2cm} {\small {\bf To appear in:} Proc.\ of the 13th North
American Workshop on CVs, \\ Jackson Hole, Wyoming, June 1997, \\eds.\
S.~Howell, E.~Kuulkers, C.~Woodward, ASP Conf.\ Series \\}
\vspace{1cm} 
CV and LMXB Population Synthesis}
\author{Ulrich Kolb}
%\author{Ulrich Kolb\altaffilmark{1}}
\affil{Astronomy Group, University of Leicester, Leicester LE1 7RH,
U.K.}
%\affil{and}
\affil{Max--Planck--Institut f\"ur Astrophysik,
Karl--Schwarzschild--Str.~1, 85748~Garching, Germany}
%\altaffiltext{1}{also: Max--Planck--Institut f\"ur Astrophysik,
%Karl--Schwarzschild--Str.~1, 85748 Garching, Germany}

\begin{abstract}
I discuss population synthesis methods in the context of low-mass
compact binaries. Two examples, both constraining
the largely unknown strength of orbital angular momentum losses, 
illustrate the power and problems of such studies.  
For CVs, the ``standard'' disrupted magnetic braking model predicts
that systems below the period gap are on average older than systems
above the gap. The corresponding difference in the space velocity
dispersion is testable by observations, independent of
brightness--dependent selection effects. 
For LMXBs, the fraction of transients among short--period neutron star
systems turns out to be an important diagnostic quantity constraining
not only the angular momentum loss rate but also the kick velocity
imparted to the neutron star at birth and the common envelope
efficiency. Small kicks ($\la 100$~km/s), low efficiencies and weak
magnetic braking are strongly favoured. 
\end{abstract}

%\keywords{Brevity,models}

\section{Introduction}

A binary population synthesis study considers global properties of a
certain binary class and tries to relate these to known (or assumed)
global properties of the progenitor population, ideally ZAMS binaries,
by following evolutionary channels forming binaries of this class.
Evolutionary timescales are usually much too long to give rise to
directly observable changes of binary parameters in a given
system. The only way to test evolutionary theories against
observations is to consider a large sample of binaries in a class at   
different evolutionary states and compare the observed properties of
this sample with results from population synthesis calculations. 

So, not surprisingly, it has become increasingly popular
to supplement the presentation of observational results with 
``predictions'' from binary synthesis studies, 
often in ignorance of the significant uncertainties 
involved. A population synthesis necessarily deals with a
large number of parameters, rendering predicted absolute numbers in
many cases meaningless. The strength of population synthesis studies
lies in the differential comparison between suggested models, allowing
one to test their sensitivity to different parameters. A further severe 
problem is the fact that selection effects may distort
the observed picture of a binary class so much that it
bears no resemblance to the true intrinsic population. 

In the following I illustrate the power and problems of population
synthesis models with two applications to low--mass compact binaries.
These are either cataclysmic variables (CVs) with a white
dwarf accretor, or low--mass X-ray binaries (LMXBs), with a neutron star
or black hole accretor. The Roche--lobe--filling donor is a low--mass
main--sequence or giant branch star. These binaries represent a
long--lived, 
interacting species at the endpoint of a chain of complex progenitor
phases, ideal for study with population synthesis methods.

\section{The General Framework}

The standard evolutionary channel leading to the formation of
low--mass compact binaries involves a common envelope phase (CE) where
the more massive star, the progenitor of the compact star, swells to
giant dimensions, overflows its Roche lobe and engulfs the 
secondary. Orbital energy, extracted by dynamical
friction, expels the envelope. If the initial orbit was
wide enough the remnant core and the secondary survive the CE as
a binary with a much smaller separation (see e.g.\ Iben \& Livio 1993 
for a review).  

The ZAMS mass $M_{\rm p}$ of the giant determines the nature of the later
compact binary. If $M_{\rm p} \la 10 \msun$ the remnant core will be 
a white dwarf, for $M_{\rm p} \ga 10 \msun$ a helium star that eventually
explodes as a supernova (SN) to leave either a neutron star (NS) or,
if the helium star is more massive, a black hole (BH). Survival
of the binary in the case of a SN depends critically on the amount of
ejected mass in the 
explosion and on kick velocity imparted to the
collapsed object if the explosion is asymmetric. The smallest $M_{\rm p}$ 
which leads to BH rather than NS formation is uncertain (e.g.\
Portegies Zwart et al.\ 1997), but usually taken to be of order
$30-40\msun$.  

The detached post--CE evolution and the subsequent (stable)
semi--detached evolution is driven either by orbital angular momentum
losses (``j--driven'') such as gravitational radiation and
magnetic braking (MB), or by the nuclear expansion of the secondary
(``n--driven''). Essentially all systems with an initial secondary
mass above $\simeq 1.5 \msun$ are n--driven. This is because the
secondary's nuclear time $t_{\rm MS}$ decreases with stellar mass
$M_2$ roughly as $t_{\rm MS} \propto M_2^{-3}$, and angular momentum
losses become inefficient for large $M_2$. 
An exception are those systems with large mass ratios $M_2/M_1$ which
are unstable to thermal--timescale mass transfer (e.g.\ supersoft
sources), which therefore maintain mass transfer without driving.

\section{Example I: The Age of CVs}

Population synthesis calculations for CVs within the above framework
have been performed by a number of authors (e.g.\ Politano 1996;
de~Kool 1992; Kolb 1993; Yungelson et al.\ 1997, and references
therein; for a review see Kolb 1995, 1996). 
Most studies aim at an interpretation of the 
deficiency of systems with orbital period between 2 and 3~h, the
``period gap'' in the orbital period distribution.
In the standard model for CV evolution (cf.\ King, this volume) the
gap arises because systems are detached and do not appear as CVs as
they evolve from 3 to 2~h. The systems are thought to
detach in response to a sudden drop of the angular momentum loss rate
(by a factor $\ga 5$). This could for example occur if magnetic
braking (MB),
the dominant driving above the gap, ceases to be effective once the
secondary becomes fully convective. Below the gap the much weaker
gravitational wave emission is likely to be the only driving mechanism.
The disparity between the evolutionary timescales above and
below the gap is reflected in the mean mass transfer rate, hence
accretion rate and mean luminosity: systems above the gap are brighter
and evolve faster. Full population models show that
the vast majority of the intrinsic CV population has evolved past the
period gap 
(or already formed below the gap). Only $\simeq 1\%$ of
all CVs actually populate the period regime above the gap at the
present Galactic epoch. However, in the observed sample this population 
imbalance is hidden because the greater brightness of the small
fraction of systems above the gap strongly increases the detection
probability. Modelling this observational selection is  
difficult, but the main effect is easily illustrated by a
visual magnitude--limited sample which to first order mimics the
actual observed sample. The resulting detection probability is
almost independent of the transfer rate, and the sample has almost equal
numbers of systems above and below the gap
(e.g.\ Kolb 1996). We conclude that the observed ratio of the number of CVs
above to the number below the gap does not test the difference in
evolutionary timescales. The ratio is primarily determined by how many
CVs already {\em form} with short or long orbital periods. 
The timescale difference is not testable by determining the absolute
magnitudes of individual CVs either
because the instantaneous mass transfer rate can deviate substantially
from the secular mean (cf.\ King, this volume). 

Recently, a viable test which is largely independent of
brightness--dependent selection effects was proposed by Kolb \& Stehle
(1996), stimulated by the compilation of CV $\gamma$--velocities by
van~Paradijs et al.\ (1996).  

It is well known and well understood from basic principles of
stellar structure and evolution that the long--term evolution of CVs
proceeds in a uniform way, such that for example the mean mass
transfer rate at a 
given period is the same for all systems (with some scatter due to
different white dwarf masses), whatever the initial period
(Stehle et al.\ 1996). 
This uniformity means that the luminosity, and therefore the
detectability, of a system is independent of its age.
On the other hand, the evolutionary timescale gives a clear
signature of the age: systems above the gap evolve much faster and
must on average be much younger than systems below the gap. The
calculation of the detailed age distribution is a typical 
population synthesis application (for details see Kolb \& Stehle
1996). Most systems above the gap are found to be younger than
1.5~Gyr, whereas systems below the gap have an average age of $3-4$~Gyr.
Note that ``age'' means the time elapsed since formation of
the ZAMS progenitor binary. Although CVs born at long periods
typically evolve into the gap in less than $10^8$~yr after turn--on of
mass transfer as a CV, there is no contradiction in having a CV age of
several Gyr above the gap; this simply means that the system spent most of its
time ``dormant'' as a detached pre--CV.  

In principle, the age distribution can be determined observationally
via (1) the cooling age of the white dwarf, or 
(2) the disperion of space ($\gamma$-) velocities.
It is difficult to deduce the mean cooling age from the
surface temperature of a WD, which is constantly heated by 
accretion, dwarf nova and nova eruptions (cf.\ Sion, this volume).
Method (2) makes use of the relation between the age of a stellar
ensemble and its space velocity dispersion, found empirically for  
nearby disc stars (e.g.\ Wielen 1977, Wielen et al.\ 1992, Freeman
1993). Although the precise shape of this relation is disputed, the
controversy centers on very old stars, a regime unimportant for our 
purposes. Adopting the assumption that CVs and CV progenitors obey the
same age--velocity dispersion relation as single disc stars leaves
option (2) as the more promising one.   

Convolving the age distribution and the age--velocity dispersion
relation yields the dispersion $\sigma_\gamma$ of 
$\gamma$-veloicities as a function of orbital period. The standard
model predicts a significant difference between $\sigma_\gamma$ for
systems above the period gap (typically $\simeq 15$~km/s) and below
the period gap ($\simeq 30$~km/s).
However, the observed sample compiled by van~Paradijs et al.\ (1996)
does not confirm such a difference. 

There are two possible resolutions of this disagreement.
First, the intrinsic scatter of the observed sample might be larger than 
estimated. More accurate radial velocity data should clarify this.
Preliminary observations dedicated to high--precision
measurements of $\gamma$-velocities are already under way (Marsh et
al.\ 1997). Second, the evolutionary theory, or some part of it, could
be wrong. An obvious way to reduce the age
difference between systems above and below the gap --- while keeping
the model for the period gap intact --- is to assume that magnetic
braking is much less effective at long orbital periods, while it has
the ``standard'' strength for periods close to 3hr, i.e.\ is
sufficient to drive the mass transfer rate $\simeq 10^{-9} \msun$/yr
needed to cause a detached phase of the right width in period.
As an example, in population models with the extreme assumption that
spin--orbit coupling is too weak to enforce corotation in detached 
systems, so that MB does not operate in the pre--CV phase,
$\sigma_\gamma$ increases overall and 
the difference above and below the gap decreases to $<
10$~km/s.

\section{Example II: The transient fraction in short--period
neutron star LMXBs}

The formation of neutron star LMXBs (NS-LMXBs) within the standard
framework has been the focus of a number of population synthesis studies
(Iben, Tutukov \& Yungelson 1996; Terman, Taam \& Savage 1996;
Kalogera \& Webbink 1996, 1997). However only  
recently has it become clear that the fraction of transients
among short--period NS-LMXBs represents an important
constraint for these models. The interpretation of soft X--ray
transient outbursts as disc instability phenomena in analogy to dwarf
nova outbursts of CVs places an upper limit on the mass
transfer rate $\dot M$ in transients. If $\dot M$ is too large the
disc is hot, hydrogen fully ionized even in the outer disc, and no
instability occurs. 
For LMXBs this critical rate is significantly smaller than for
CVs because in the former the strong irradiation from the central
source heats the disc and keeps it stable even for a relatively small
transfer (accretion) rate (van~Paradijs 1996; King, Kolb \& Burderi
1996; see also King, this volume). The effect is particularly marked
for neutron star systems, and less severe for black hole systems
(King, Kolb \& Szuszkiewicz 1997). 

The secular mean mass transfer rate in most n--driven LMXBs and all
black hole LMXBs is low enough for disc instabilities to occur, while
all j--driven NS-LMXBs with ZAMS donors would be
persistently bright if magnetic braking (MB) had the same functional
form as for CVs, with a strength tightly constrained 
at 3~h orbital period from the width of the CV period gap. The
corresponding transfer rate is so much larger than the critical rate
for transient behaviour in 
NS-LMXBs that any reasonable ``sub--standard'' strength magnetic
braking at longer periods would barely allow transients even
at $P \ga 10$~h. 

However, from observations it is clear that the fraction $r$ of
transients 
among short--period NS-LMXBs is non--negligible, although it is
difficult to estimate the true {\em  intrinsic} value of $r$. 
Three out of 15 NS (or suspected NS) systems with periods in the range
$3 < P/{\rm h} < 18$ are classified as transient (Ritter \& Kolb 1997).
Most persistent sources in the Galaxy in this period range should
have been detected by now, and most of the $\simeq 30$ persistent sources
with undetermined orbital period are probably short--period
systems. Hence a fairly pessimistic lower limit for $r$ is
$5\%$. It is very likely that there are many more so far undetected
short--period transients, perhaps more than 100, 
and that some of the $\simeq 30$ known transients with undetermined
periods are also short--period NS systems. 
The corresponding rather large value of $r$ implies
that the mass transfer rate in a large fraction of j--driven
NS-LMXBs should be much smaller than the secular mean transfer rate in
CVs. As fluctuations of the instantaneous mass transfer rate around the
secular mean are probably unimportant in j--driven LMXBs (cf.\ King, this
volume), the large transient fraction demands a small secular
mean $\dot M$ in these systems. The evolutionary state of the donor
offers a promising explanation for this. 

It is well--known that in j--driven evolution with a 
Skumanich--type magnetic braking formalism ($\dot J \propto \Theta
R^\rho \omega^3$, where $\Theta$ and $R$ is the secondary's moment of
inertia and radius, $\omega$ the binary angular velocity, and typically
$rho \simeq 2-4$)
the mass transfer rate is smaller if the donor is somewhat
nuclear--evolved (e.g.\ Pylyser \& Savonije 1988; Singer et al.\
1993). ``Somewhat'' means that the secondary is still in the core
hydrogen burning phase, but close to its end.  
To quantify this we define $f=t_{\rm turn-on}/t_{\rm MS}$, the age
$t_{\rm turn-on}$ of the binary at turn--on of mass transfer in units  
of the secondary's main--sequence lifetime $t_{\rm MS}$.
Then j--driven systems correspond to $f<1$, and $\dot M$ is 
smaller the closer $f$ is to unity. The reason is twofold:
first, an evolved secondary is ``undermassive'', i.e.\ has less mass
than a ZAMS star filling its Roche lobe at the same period would
have. This in turn implies a smaller angular momentum loss rate as
$\dot J \propto M_2^{\rho/3}$ with $R \propto M_2^{1/3}$ from Roche
geometry. 
Second, evolved stars are more centrally condensed, hence the
moment of inertia is smaller. 

For simplicity we identify the fraction $r$ of transients with the
fraction $r(f>f_0)$ of those systems in the population which have a
secondary that is more evolved than $f_0$. 
Until detailed calculations of the secular evolution for LMXBs
for a wider range of initial secondary masses, MB strengths and values
of $f$ are performed the critical value $f_0$ allowing
transient behaviour must be treated as a free parameter. We expect
$f_0 \ga 0.8 - 0.9$ (e.g.\ Pylyser \& Savonije 1988).

King \& Kolb (1997) investigated the formation of LMXBs under the
assumption of spherically symmetrical SN explosions and 
magnetic braking following the prescription of Verbunt \& Zwaan
(1981). They showed that 
j--driven NS-LMXBs can form only under very special formation
conditions (cf.\ King, this volume) which require that the secondaries
are fairly massive ($\ga 1.2 \msun$) at mass transfer turn--on, and
that indeed a large fraction of systems have $f>f_0$. 

However, this fraction is smaller if the SN is asymmetric and the
neutron star receives a kick velocity. Two effects of kicks
matter most. First, binaries which would have been
unbound by a symmetric SN, i.e.\ systems with relatively small
secondary masses $M_2$, can survive the explosion if the kick is
suitably directed. Hence the mean
secondary mass in the post--SN binaries decreases with increasing
kick speed.
Second, in the presence of kicks the post--SN orbital separation after
circularisation is on average smaller, so that the systems spend less
time in the detached state. Both effects decrease the mean 
value of $f$ in the population. 

Evidence for the presence of significant kick velocities at the
formation of neutron stars is substantial (e.g.\ Kaspi et al.\ 1996,
Hansen \& Phinney 1997, Lorimer et al.\ 1997, Fryer \& Kalogera 1997),
and the question arises  
if it is possible to form j--driven NS-LMXBs with moderate kick
velocities but still maintain a large fraction $r(f>f_0)$. A natural
way to ``compensate'' the reduction of $r$ from kicks is to reduce the
MB strength at longer periods. This keeps the systems longer in the
detached post--SN state and allows the secondary to evolve. 

Again, testing this is a typical application for population synthesis
calculations. Below I present some results of a systematic
study which makes use of the models by Kalogera \& Webbink (1996,
1997). Details will be presented elsewhere (Kalogera, Kolb \& King
1997).  

Figure~1 shows the transient fraction $r(f>f_0)$ as a function of $f_0$
for various parameter combinations. The upper panels assume 
inefficient CE ejection (efficiency parameter $\alpha_{\rm CE} = 0.3$,
i.e.\ only $30\%$ of the released orbital energy is used to unbind the
envelope), and the lower panel efficient ejection ($\alpha_{\rm CE} =
1.0$). The kick velocity distribution is assumed to be Maxwellian,
with a mean value $<v^2_{\rm kick}>^{1/2}$ of 20~km/s (left), 100 km/s
(middle) and 300~km/s (right). The linestyle indicates the magnetic
braking law. Solid lines correspond to ``strong'' braking (as in King \&
Kolb 1997), dashed lines to ``weak'' braking (as in Kalogera \& Webbink
1997). Both MB descriptions are Skumanich--type laws, calibrated to
the CV period gap, with $\rho=4$ and $2$ in the strong and weak
case. In addition, the weak case 
involves a $M_2$--dependent reduction factor which drops exponentially
for $M_2>1.0\msun$. 

%\plotfiddle{fig2.ps}{8cm}{00}{50}{50}{-200}{-30} 

\begin{figure}[t]
\begin{center}
\unitlength1cm
\begin{picture}(13,9,2)
\thicklines
\put(2.5,0.2){\vector(1,0){7}}
\put(9.8,0.1){mean kick velocity}
\put(0.3,8.5){\vector(0,-1){6.5}}
\put(0,9.2){CE} 
\put(0,8.8){efficiency}
%\put(1,1){\plotfiddle{fig.ps}{8cm}{00}{45}{45}{-400}{-30}} 
\put(-12.5,0.3){\plotfiddle{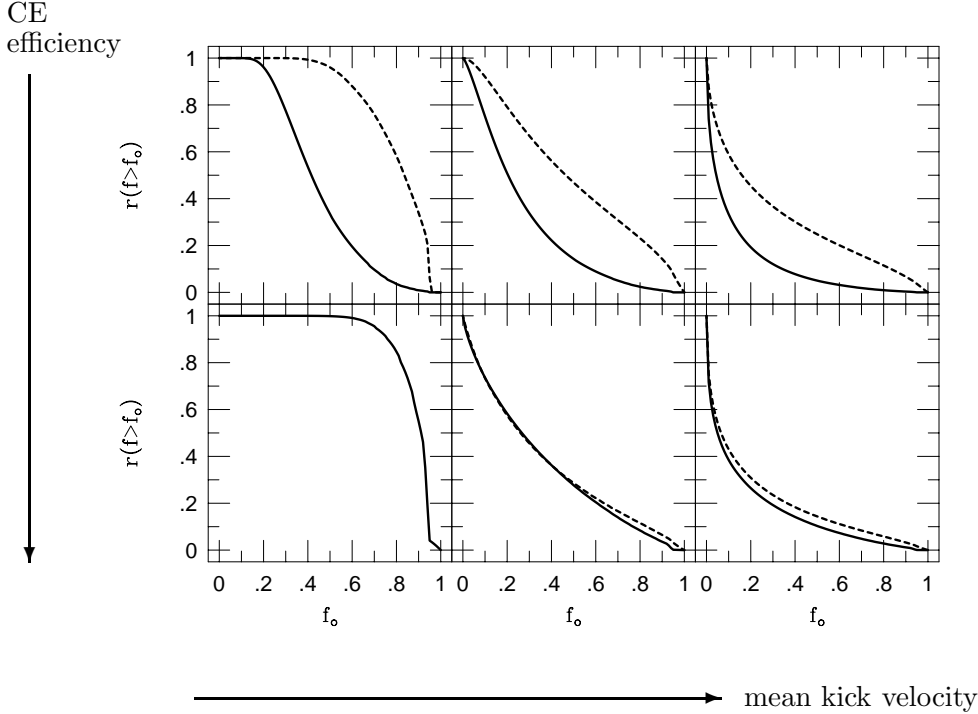}{8cm}{00}{45}{45}{0}{0}} 
\end{picture}

\caption{\small The transient fraction $r(f>f_0)$ in
short--period neutron 
star LMXBs as a function of the critical evolutionary state $f_0$ for
transient behaviour, for different model parameters. Full line: strong
magnetic braking; dashed line: weak magnetic braking. Upper panels:
$\alpha_{\rm CE}=0.3$, lower panels: $\alpha_{\rm CE}=1.0$. Left
panels: mean kick velocity 20 km/s; middle panels: 100 km/s; right
panels 300 km/s. See text for further details.
}

\end{center}
\end{figure}

%This is to account for the absence of rotational
%braking in more massive stars with radiative (instead of convective)
%envelopes. 

The strong MB case with small kick and efficient CE ejection 
shown in the lower left panel confirms the result of
King \& Kolb (1997) for spherical symmetrical SNe: a large fraction
of systems are close to $f_0=1$. 
In the corresponding model with weak MB, j--driven systems do not form
in significant numbers, and the curve $r(f_0)$ is not shown in this panel.
This is due to constraints on the primary's progenitor evolution.
Roche lobe overflow must occur after core He burning
but prior to the SN explosion. As the radius growth of massive stars
in this phase is rather limited (Schaller et al.\ 1992) the allowed
range of initial orbital separations is very narrow. Therefore the
range of post--CE separations is narrow as well. 
If the CE ejection is efficient ($\alpha_{\rm CE}$
large) the orbits in this narrow range are 
generally too wide for the weak angular momentum losses to
establish contact within a Hubble time (cf.\ Kalogera \& Webbink
1997).  

As expected, with increasing mean kick velocity the fraction of
unevolved systems (small $f$) in the population increases
significantly, while the transient fraction ($f \ga 0.8$) decreases
sharply. Because of the restricted range of post--CE orbits the effect of
changing $\alpha_{\rm CE}$ is more complex. The curves $r(f_0)$ are
insensitive to the MB strength for efficient CE ejection, while 
for inefficient CE ejection the transient fraction is generally larger
in the case of weak MB. Comparing curves for the same MB law and 
$<v^2_{\rm kick}>^{1/2}$ but different $\alpha_{\rm CE}$ shows that in
the weak MB case the transient fraction is larger for $\alpha_{\rm
CE}=0.3$ whereas in the strong MB case it is larger for $\alpha_{\rm
CE}=1$.

With our estimates $f_0 \simeq 0.8$ and $r \ga 0.3$, and assuming that
the kicks are not generally negligible, the models clearly favour
small kick velocities, inefficient CE ejection and weak MB. In
particular, it is hard to reconcile the necessity for a significant
number of short--period transient NS-LMXBs with a mean kick velocity
in excess of 100~km/s.  

An important additional constraint on these parameters comes from the
ratio of long--period (n--driven) to short--period (j--driven)
NS-LMXBs. If the kicks are too small and MB too weak, too many
n--driven systems would form, contrary to what is observed.

Taken together, tight constraints can be placed on all three
parameters.

\acknowledgments
I thank Vicky Kalogera for very useful discussions and for preparing
Figure~1. I am grateful to Andrew King for comments and for improving
the language of the manuscript and to Rudolf Stehle for a careful
reading of the text.


\begin{references}

{\small

\reference Freeman K.C. 1993, in Galaxy evolution: The Milky Way
Perspective, ed.\ S.R.~Majewski, ASP Conf.\ Ser.\ 49, (SanFrancisco:
ASP), 125  
\reference Fryer C., \& Kalogera V. 1997, \apj, in press
\reference Hansen B.M.S., \& Phinney E.S. 1997, \mnras, in press
\reference Iben I., \& Livio M. 1993, \pasp, 105, 1373
\reference Iben I., Tutukov A.V., \& Yungelson I.R. 1995, \apjsupp,
100, 233
%\reference Kalogera V. 1997, in ...
%\reference Kalogera V. 1996, \apj, 471, 352
\reference Kalogera V., \& Webbink R. 1997, \apj, in press
\reference Kalogera V., \& Webbink R. 1996, \apj, 458, 301
\reference Kalogera V., Kolb U., \& King A.R. 1997, in preparation
\reference Kaspi V.M., Bailes M., Manchester R.N., Stappers B.W., \&
Bell J.F. 1996, Nature, 381, 584
\reference King A.R., this volume
\reference King A.R., \& Kolb U. 1997, \apj, 481, 918 
\reference King A.R., Kolb U., \& Burderi L. 1996, \apj, 464, L127
\reference King A.R., Kolb U., \& Szuszkiewicz E. 1997, \apj, in press
(Oct 10 issue, vol. 488)   
\reference Kolb U. 1996, in Cataclysmic Variables and
 Related Objects, ed.\ A.~Evans, J.H.~Wood, IAU Coll.~158, (Dordrecht:
Kluwer), 433 
\reference Kolb U. 1995, in Cape Workshop on Magnetic Cataclysmic 
 Variables,  ed.\ D.A.H.~Buckley, B.~Warner
 ASP Conf.\ series Vol.~85, (San Francisco: ASP), 440
\reference Kolb U., \& Stehle R. 1996, \mnras, 282, 1454
\reference Kolb U. 1993, A\&A, 271, 149 
\reference de~Kool M. 1992, \astap, 261, 188 
%\reference Lyne A.G., \& Lorimer D.R. 1994, Nature, 369, 127
\reference Lorimer D.R., Bailes M., \& Harrison P.A. 1997, \mnras, in press 
\reference Politano M. 1996, \apj, 465, 338
\reference Portegies Zwart S.F., Verbunt F., \& Ergma E. 1997, \astap,
321, 207
%\reference Pylyser \& Savonije 1988a, \astap, 191, 57
\reference Pylyser \& Savonije 1988, \astap, 208, 52
%\reference Ritter H. 1996, in Evolutionary Processes in Binary Stars,
%ed.\ R.A.M.J.~Wijers, M.B.~Davies, C.A.~Tout, NATO ASI Series C,
%Vol.~477, (Dordrecht: Kluwer), 223
\reference Ritter H., \& Kolb U. 1997, A\&AS, submitted
\reference Schaller G., Schaerer D., Meynet G., \& Maeder A. 1992,
A\&AS, 96, 269
\reference Singer R., Kolb U., \& Ritter H. 1993, AG Abstract Series,
9, 39
\reference Sion E., this volume
\reference Stehle R., Ritter H., \& Kolb U. 1996, \mnras, 279, 581
\reference Terman J.L., Taam R.E., \& Savage C.O. 1996, \mnras, 281,
552 
\reference Van~Paradijs J. 1996, \apj, 464, L139
\reference Van~Paradijs J., Augusteijn T., \& Stehle R. 1996, A\&A,
312, 93
\reference Verbunt F., \& Zwaan C. 1981, \astap, 100, L7
\reference Wielen R.\ 1977, \astap, 60, 263
\reference Wielen R., et al. 1992, in The Stellar Populations of
Galaxies, ed.\ B.~Barbuy, A.~Renzini, IAU Symp.~149, (Dordrecht:
Kluwer), 81 
\reference
Yungelson L., Livio M., \& Tutukov A. 1997, \apj, 481, 127

}

\end{references}
\end{document}